# UNE APPROCHE ORIENTEE OBJET DE LA PROTECTION DANS LES SYSTEMES D'INFORMATION.

Joël COLLOC

Laboratoire d'Informatique Médicale
des Hospices Civils de Lyon.
162, Avenue Lacassagne
69424 Lyon Cedex 03
Tél : 72.40.70.86

Laboratoire d'Informatique
et Automatique Appliquées.
INSA de Lyon
20, Avenue Albert Einstein
69621 Villeurbanne Cedex.

RESUME : Le système de protection présenté exploite l'encapsulation, la communication par messages, les fonctions d'interface d'un modèle orienté objet décrit dans de précédents travaux. Chaque personne est représentée au système par son objet "UTILISATEUR". La procédure de reconnaissance est adaptable au gré de chacun. Les objets et types d'un utilisateur portent sa signature personnelle, fournie et connue seulement par le système. Les droits de l'administrateur sont réduits aux sauvegardes. Le contrôle des accès concerne chaque message, le système est solide car cloisonné, souple, adaptable et psychologiquement acceptable.

Mots-clés : Système de protection, modèle orienté objet, objet utilisateur, signature du propriétaire, encapsulation, messages

ABSTRACT : We provide a protection system making use of encapsulation, messages communication, interface functions coming from an object oriented model described in previous works. Each user represents himself to the system by the mean of his "USER" object type. The recognition procedure is suitable to every one's needs. Any user's objects and types are labeled with a personal signature, exclusively provided and known by the system. Administrator's rights are restricted to backup procedures. The system verify each messages access, it is robust because partitioned, flexible, suitable and psychologically acceptable.

1. INTRODUCTION

Les systèmes informatiques contiennent très souvent des informations essentielles et secrètes dont la sécurité revêt deux aspects complémentaires : Assurer l'intégrité des données et se prémunir contre les accès, aux ressources de la machine, par des personnes non-autorisées. Nous rappelons quelques mécanismes de prévention, couramment employés dans divers systèmes d'exploitation.

Nous proposons une approche innovante de la protection, exploitant les concepts d'un modèle orienté objet [COLL 89] [COLL 90] : chaque personne est représentée dans le système par une instance du type d'objet "Utilisateur". Les objets nouvellement créés reçoivent la signature de leur propriétaire. L'encapsulation, notamment la partie privée des objets, est exploitée pour dissimuler les données et les fonctions aux personnes non-autorisées. L'acceptation ou le refus des requêtes a lieu grâce à des messages de contrôle lancés par le système à l'insu de l'utilisateur. Tous les utilisateurs disposent des mêmes moyens de protection; l'administrateur du système ne dispose que de pouvoirs de sauvegarde sur les objets de certains utilisateurs perdus de vue.

2. L'ETAT DE L'ART EN MATIERE DE PROTECTION

Les problèmes de protection sont à la fois d'ordre technique et psychologique.

   2.1. Les buts du système de protection

Le but du système de protection est d'interdire l'accès aux données et l'exploitation des programmes, par des personnes non-autorisées.

   2.2. Les motivations des intrusions dans un système informatique.

Les motivations pour pénétrer indûment dans un système informatique comprennent schématiquement : la simple curiosité, le défi et l'excitation (anime surtout les "hackers")[BURG 89], la vengeance, la malveillance (conflit professionnel...), l'escroquerie et le banditisme (extorsion de fonds, d'avantages au détriment des banques ou des sociétés), le recueil de renseignements concernant une personne, l'espionnage industriel ou militaire dans les domaines de haute technologie...

   2.3. Quelques techniques utilisées pour entrer par effraction dans un système informatique.

La stratégie du cheval de Troie utilise une procédure, espionnant les utilisateurs, greffée au sein d'un programme qui conserve, par ailleurs, un fonctionnement apparemment normal.

294                                   VERSION SOUMISE - INFORSID'91 - PARIS PANTHEON-SORBONNE

Les programmes résidents, chargés au démarrage, sont protégés durant la session par le système. Ils redirigent les vecteurs d'interruptions, détournent, les appels aux primitives du noyau : interceptent les caractères frappés au clavier, les lectures et écritures sur les disques et les ports d'entrée/sortie. Les programmes résidents chaînent enfin vers la routine d'origine qui effectue le travail prévu et masque ainsi la supercherie. [HYMA 86] Il devient facile d'intercepter les mots de passe, fournis par les utilisateurs, lors de leur connexion. Une variante consiste à afficher un faux message d'invite (genre login:). [STOL 89]

Les "virus" informatiques sont des programmes parfois résidents qui possèdent en outre la propriété de se reproduire et de répandre des descendants dans les systèmes et les réseaux. Un virus peut détruire des fichiers de données ou de code ou bien jouer le rôle de cheval de Troie [BURG 89].

L'exploitation de failles des systèmes d'exploitation :

Le maintient en mémoire centrale de données logiquement détruites autorise leur consultation et leur sauvegarde. La modification directe des attributs des fichiers sur le disque.

Sous UNIX(R) l'exécution momentanée de certains processus en mode noyau permet d'acquérir indûment des privilèges étendus. [TANE 89] [STOL 89]

### 2.4. Les critères architecturaux du système de protection

Il est vain de tenter de protéger des applications si le système d'exploitation n'est pas sûr. Tout ce qui est réalisé pour assurer la protection peut être contourné.

En effet, on ne peut pas empêcher l'étude du code du système d'exploitation. Toutefois, on peut rendre cette analyse difficile, les moyens dont dispose le concepteur étant la complexité et le temps.

Une architecture hiérarchisée de système d'exploitation est plus fragile qu'une structure non-hiérarchisée : l'extorsion du statut de super-utilisateur procure, à un intrus, tous les privilèges et la maîtrise totale du système. Une architecture non-hiérarchisée est cloisonnée comme un sous-marin, l'effraction d'un "compte utilisateur" demeure limitée. L'existence d'une seule primitive de connexion rend le système plus vulnérable car la reconnaissance de l'utilisateur est concentrée en un seul point.

En 1975, Saltzer et Schroeder énoncent six principes [TANE 89] 1. Il faut rendre la conception du système publique. 2. Le moindre doute doit faire refuser l'accès aux objets. 3. Il faut vérifier les droits d'accès régulièrement. 4. Il faut donner à chaque processus le moins de privilèges possible. 5. Le mécanisme de protection doit être simple, uniforme, implanté dans les couches basses du système. 6. Le mécanisme doit être psychologiquement acceptable.

### 2.5. Les aspects psychologiques du système de protection.

Cet aspect concerne l'identification des utilisateurs par une ou plusieurs de leurs caractéristiques. Les contraintes imposées aux utilisateurs doivent être bien tolérées, sous-peine de non-utilisation, rendant illusoire les efforts techniques consentis. En général, la reconnaissance d'un utilisateur a lieu grâce à un mot de passe, réclamé au moment de la connexion. L'utilisateur doit le mémoriser et ne pas l'écrire. Il doit en changer souvent. Les mots de passe figurent sous forme chiffrée dans le système. Ils ne comportent jamais d'information concernant directement les utilisateurs. Le choix hors du dictionnaire interdit une recherche systématique à l'aide de lexiques informatisés. [STOL 89]

Les dispositifs d'identification physique demeurent souvent coûteux et mal acceptés par les utilisateurs en dehors de milieux classés secrets.

L'excès de privilèges de l'administrateur dans les systèmes hiérarchisés comme UNIX (R) pose parfois des problèmes psychologiques aux utilisateurs.

## 3. RAPPEL DES CONCEPTS DU MODELE ORIENTE OBJET UTILISE

Le modèle orienté objet utilisé a été présenté dans [COLL 89] et [COLL 90], nous le rappelons succinctement ici.

### 3.1. Le modèle établit un double niveau conceptuel

Un niveau intérieur nommé endobjet regroupant d'une part les relations de composition de sous-objets (est_partie_de) et les relations regroupant les attributs concernant un objet (a_un). Les relations de composition établissent un héritage intérieur multiple ascendant transmettant des propriétés des sous-objets composants vers l'objet composé. Le niveau intérieur encapsule des attributs et des fonctions représentant les caractéristiques statiques et le comportement des objets.(voir figure 1)

Un niveau extérieur, nommé exobjet, représente l'environnement d'un objet, où résident les relations de généralisation /spécialisation (est_un) unissant les types d'objet aux sous-types d'objet. Ces relations établissent une hiérarchie de types et un héritage simple descendant des caractéristiques d'un type d'objet vers ses sous-types. Les niveaux intérieur et extérieur sont reliés par l'instanciation qui produit de nouveaux objets conformes à chaque type. L'évolution spontanée de l'état des objets est exprimée par des fonctions dynamiques de l'endobjet, modifiant automatiquement, dans le temps, les liens de composition et la valeur de référence des attributs.

Des fonctions d'évaluation de l'endobjet comparent la structure d'un objet (ses sous-objets composants), la valeur de certains attributs avec celles d'autres objets.[COLL 87]

La communication entre objets a lieu à l'aide de messages et de fonctions d'interface.

### 3.2. L'encapsulation et les fonctions d'interface

Les fonctions d'entrée ou d'affectation vérifient les valeurs destinées aux attributs de l'endobjet (contraintes de cardinalité et d'intégrité) puis les affectent, éventuellement sous forme chiffrée.

Les fonctions de sortie ou de consultation des attributs nous renseignent sur l'état d'un objet. Déclenchées par la réception d'un message entrant, conforme, elles accèdent à la valeur de certains attributs, les décodent si besoin, puis produisent un message de réponse sortant, porteur de la valeur des attributs concernés. Le propriétaire de l'objet choisit ainsi les attributs consultables de l'extérieur et ceux qui resteront privés, connus de lui seul.

### 3.3. Les messages.

Les messages communiquent des informations d'un objet émetteur vers un ou plusieurs objets récepteurs appelés cibles.

Chaque message précise : 1. l'identifiant de l'objet émetteur, son type et la signature de son propriétaire, 2. la définition de la ou des cibles : soit un objet déterminé par son identifiant, soit toutes les instances d'un type d'objet, rendant ainsi possible des actions génériques.

3. le nom de la fonction à déclencher désignant l'action à effectuer, au sein du ou des objets cibles, accompagné éventuellement d'arguments et de leur type. 4. La nature de la réponse au message et s'il y a lieu, le type de celle-ci. Par défaut, elle est envoyée uniquement à l'objet émetteur. Toutefois, une copie peut, à la demande, parvenir à d'autres objets ou types d'objet.

Shibayama et Yonezawa dans le langage acteur ABCL/1 proposent des messages porteurs de conditions de valeurs pour certains attributs sélectionnant les objets cibles. [SHIB 87] [YONE 86] [YONE 87] Nous préférons confier la vérification des conditions d'accès aux fonctions intérieures d'interface. [COLL 90]

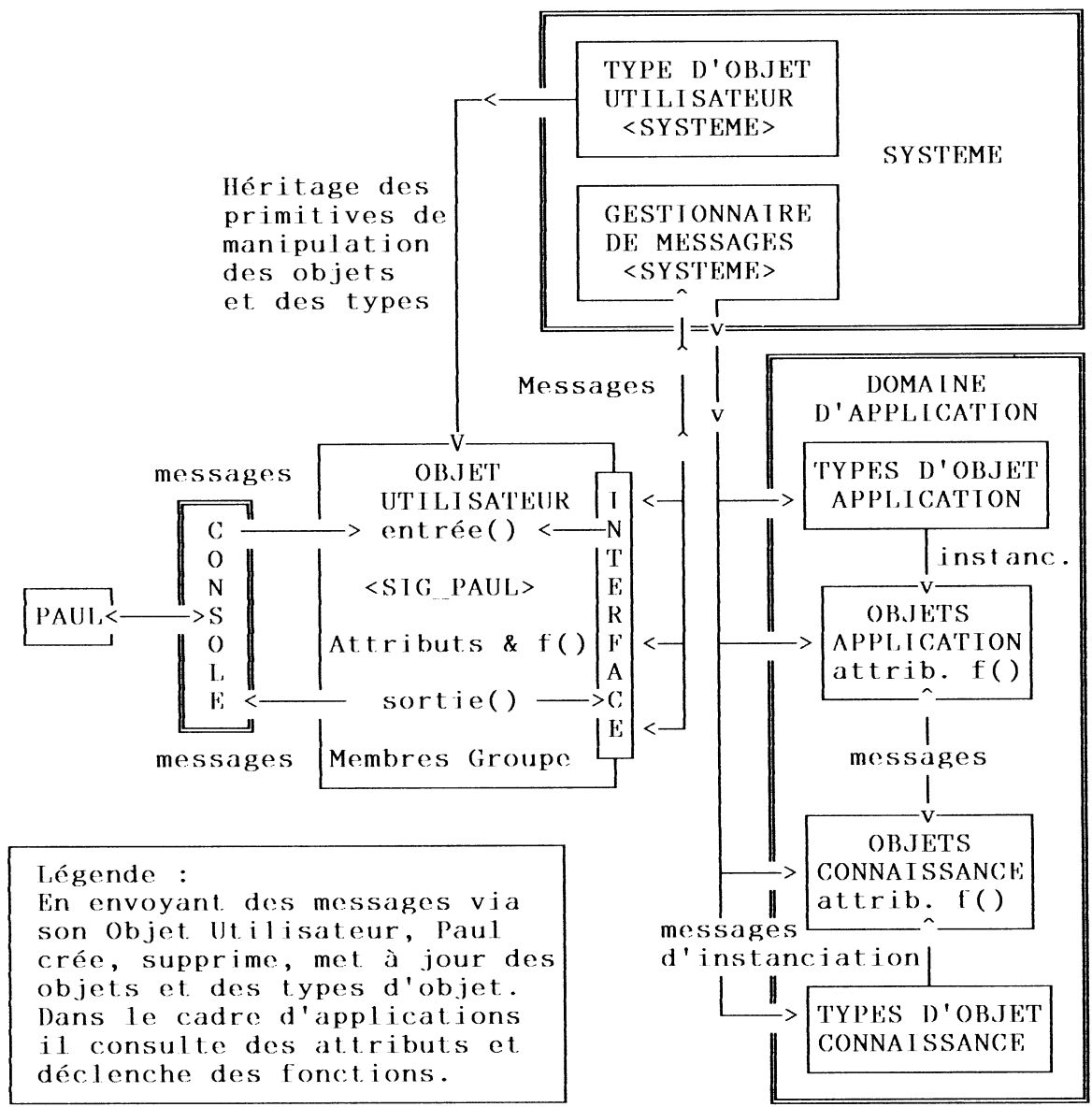

Schéma n°1 : COMMUNICATION ENTRE UN UTILISATEUR ET LE SYSTEME.

## 4. PRESENTATION DU SYSTEME DE PROTECTION

### 4.1. Le type d'objet utilisateur.

Notre approche est innovante car chaque personne accédant au système est représentée par une instance du type d'objet "UTILISATEUR" dont le système est propriétaire. Paul exécute des actions grâce aux fonctions d'entrée/sortie assurant l'interface, avec son propre objet utilisateur (Cf Schéma 1)

Par son intermédiaire, Paul lance des messages à destination d'objets ou de types d'objet pour exécuter des fonctions nécessaires à la construction, au lancement d'une application. Il peut exploiter ou consulter d'autres objets ou types d'objets pour lesquels il a reçu des droits. Il peut interroger les attributs d'objets existant, créer de nouveaux types et sous-types, instancier des objets, composer des objets complexes à partir d'objets simples, ajouter des attributs et modifier leurs contraintes, définir des fonctions intérieures. Il accorde ou retire, à d'autres utilisateurs, des droits d'accès sur les types et les objets dont il est propriétaire. Ces fonctions de gestion des objets appartiennent au noyau du système et sont donc inaccessibles.

### 4.2. La reconnaissance de l'utilisateur

Chaque utilisateur décide librement des actions qui assureront son identification par le système au moment d'une tentative de connexion. Il établit un protocole de reconnaissance en modifiant les attributs de son objet utilisateur, dont la structure doit toutefois rester conforme au type UTILISATEUR. Ce dernier prévoit, par défaut, des mesures minimales de contrôle de connexion inamovibles. L'utilisateur choisit les attributs à renseigner : nom, un mot de passe... Il peut décider, au contraire, que certains attributs ne devront pas être renseignés. L'identification peut avoir lieu, en comparant les messages de la personne tentant de se connecter avec une séquence d'actions retenues comme significative par l'utilisateur, dans un certain laps de temps. Le système reconnaît l'utilisateur grâce à des fonctions d'évaluation qui établissent si les valeurs entrées, les actions effectuées sont conformes aux habitudes représentées par les attributs encapsulés dans l'objet utilisateur. Un tel protocole met en échec toutes tentatives de recherche systématique, par des algorithmes de type essais / erreurs. En cas de reconnaissance satisfaisante, l'utilisateur accède à tous les objets dont il a la propriété et ceux pour lesquels des droits lui ont été accordés par les propriétaires respectifs.

### 4.3. L'administrateur du Système

A l'inverse des systèmes hiérarchisés comme UNIX(R), où l'administrateur possède tous les droits [TANE 89] [BOUR 85] [LUCA 86], nous ne lui accordons qu'une tâche de sauvegarde. Cette approche est plus solide car il ne suffit pas d'extorquer le statut de "super utilisateur" pour se rendre maître de tout le système [TANE 89]. L'objet particulier "administrateur" préexiste au sein du système, il reconnaît comme tel, celui qui possède le numéro de série du système et d'un mot de passe, situé sous forme chiffrée, non pas dans un fichier comme dans UNIX (R) mais quelque part dans le code du système. Cet objet administrateur dispose de droits particuliers pour assurer les sauvegardes des travaux et instancier de nouveaux objets de type "UTILISATEURS" dans le système.

Toutefois, les actions d'accès aux objets (lecture, écriture, exploitation), réservées aux utilisateurs, lui sont interdites. Afin de préserver l'équité face au système, nul ne peut se connecter à la fois comme administrateur et comme utilisateur.

Le type UTILISATEUR et l'objet administrateur sont définis par le système et ils ne peuvent être modifiés. L'intervention de l'administrateur gère une situation exceptionnelle : un utilisateur décède ou démissionne : comment récupérer son travail ? L'administrateur transfère en bloc la propriété de la totalité des types et des objets de la personne quittant le système vers un nouveau propriétaire dont ils reçoivent la signature. Cette opération revient à détruire l'objet utilisateur de la personne concernée et à l'exclure du système.

L'administrateur peut transférer la propriété des objets au profit de son propre objet utilisateur puis se connecter en tant qu'utilisateur. Mais, l'ancien propriétaire spolié de tout son travail ne peut plus accéder au système ce qui ne passe pas inaperçu. Cette procédure de sauvegarde exceptionnelle permet de récupérer le travail d'un utilisateur quittant définitivement le système. Il s'agit d'une politique du tout ou rien : si l'utilisateur existe pour le système, il a des droits, sinon il est totalement exclu.

L'administrateur ne peut rétrocéder secondairement, par donation des types et des objets sans que l'utilisateur concerné soit consentant. Il ne peut pas utiliser son statut pour transférer de manière intempestive des types ou des objets, d'un utilisateur vers un autre, sans avoir de gros problèmes relationnels avec ses collaborateurs.

## 5. ORGANISATION DES DOMAINES DE PROTECTION

Les domaines de protection établissent à tous moments les droits d'accès en lecture, écriture et utilisation sur les objets et types d'objet existant, pour chaque utilisateur du système.

### 5.1. Définition des domaines de protection

La figure 2, ci-après, montre les domaines de protection des autres utilisateurs relativement à chaque propriétaire et la manière dont il peut accorder des droits d'accès.

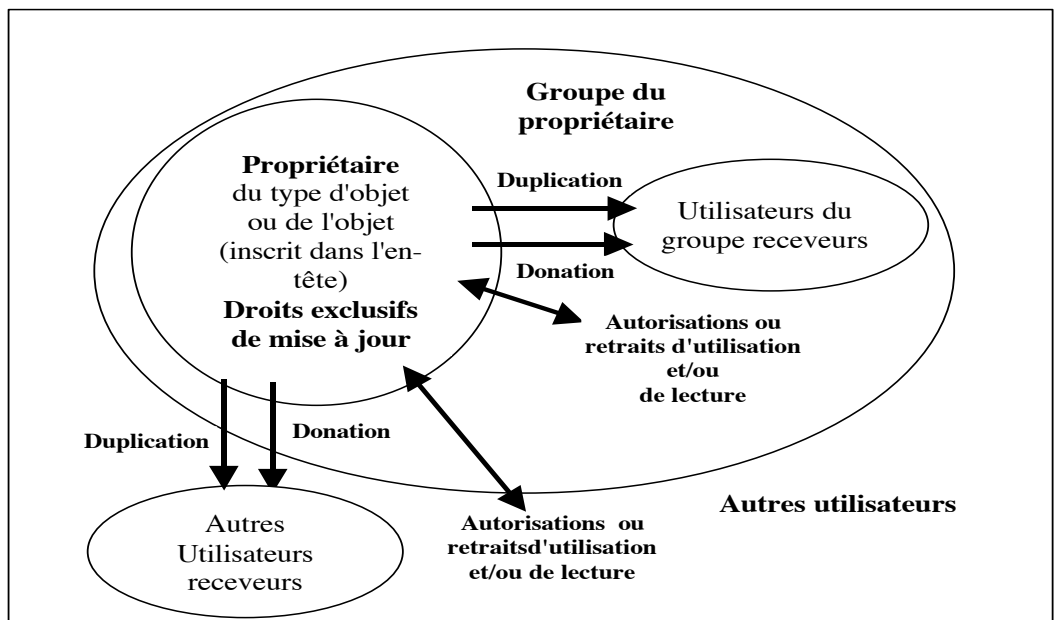

Figure 2 : les domaines de confidentialité

5.2. La notion de signature

Les objets utilisateurs sont dotés, dès leur création, d'une signature codée sur 4 octets ($2^{32}$= 4.294.967.296 combinaisons) fournie par une fonction de "hashage" (voir figure 3). Lorsque les utilisateurs créent des objets ou des types, le système les estampille avec la signature de leur propriétaire. Cet attribut particulier est privé, protégé par l'encapsulation et consulté par les primitives du système chargées des différentes transactions (gestionnaire de messages, création, suppression, mise à jour, donation, duplication...), il ne peut être lu par personne. Par la suite la signature identifie tous les messages provenant d'un utilisateur ainsi que tous les nouveaux objets et types d'objets qu'il construit. La connaissance de cette signature n'apporte rien à un intrus, car il n'a pas la possibilité d'instancier lui même des messages en dehors des primitives du système.

5.3. Les privilèges du propriétaire sur ses objets

Les notions présentées s'appliquent aussi bien aux objets qu'aux types d'objet.
1. La donation permet à un propriétaire d'abandonner tous ses droits sur un objet au profit d'un autre utilisateur devenant ainsi le nouveau propriétaire. Il peut ainsi créer une instance pour le compte d'un autre utilisateur.
2. La duplication : Le propriétaire effectue une copie conforme d'un objet dont il confie la propriété à un autre utilisateur, tout en gardant ses droits sur l'original.
3. La cession de droits d'accès à un autre utilisateur : Le propriétaire peut céder des droits de lecture ou d'utilisation soit à son groupe exclusivement, soit à l'ensemble des utilisateurs du système. Mais il est le seul à posséder le droit d'écriture qui n'est cessible que par le biais d'une duplication ou d'une donation.
Ceci garantit l'intégrité de ses objets et types d'objet. L'autorisation en lecture permet l'accès au contenu des attributs d'un objet, par le biais des fonctions interface de consultation. L'autorisation d'utilisation concerne soit le déclenchement des fonctions d'un objet, soit l'instanciation d'un type d'objet.
4. Le retrait des autorisations : à tout moment le propriétaire d'un objet peut interdire globalement l'accès de tous ses objets à un utilisateur en le retirant de son groupe au niveau de son objet utilisateur. Il peut également retirer le droit de lecture ou d'utilisation de certains objets seulement à son groupe ou à tous les utilisateurs en modifiant les bits situés dans la zone de protection des objets concernés (Cf figure 3). Le propriétaire peut enfin choisir d'interdire sélectivement la consultation de certains attributs de ses objets en modifiant les conditions d'accès au niveau des fonctions d'interface.

5.4. Le contrôle des messages accédant aux objets.

Lorsqu'un utilisateur expédie un message d'accès vers un objet, le système lit la signature du propriétaire dans l'en-tête de l'objet et la compare à celle du demandeur. Si elle est identique, c'est le propriétaire qui demande l'accès, l'exécution de l'action correspondant au message est déclenchée immédiatement dans l'objet récepteur. Cette situation est la plus fréquente car un utilisateur exploite le plus souvent ses propres objets.

Dans le cas contraire, il s'agit d'un autre utilisateur et une fonction vérifie les bits de l'attribut de protection.

Trois cas peuvent alors se produire : 1. L'accès demandé (lecture ou exécution) est autorisé pour tous les utilisateurs : aucune vérification supplémentaire n'est nécessaire et l'action correspondant au message est exécutée. 2. L'accès demandé est interdit à tous les utilisateurs autres que le propriétaire : l'action n'est pas exécutée, un message d'erreur est retourné à l'objet utilisateur requérant. 3. L'accès demandé n'est autorisé que pour les membres du groupe : un message de contrôle de statut, portant l'identifiant du demandeur, est expédié à l'objet utilisateur propriétaire (dont la signature réside dans l'objet récepteur). Au sein de l'objet utilisateur propriétaire, une fonction vérifie le statut du demandeur : membre ou non de son groupe, en consultant l'attribut privé définissant la liste des utilisateurs inscrits dans son groupe (voir figure 4). Un message de réponse confirme ou infirme l'appartenance du demandeur au groupe de l'objet propriétaire, ce qui provoque respectivement l'autorisation ou l'interdiction de l'accès, assorti d'un code d'erreur.

Une méthode de contrôle astucieuse consiste à comptabiliser dans un attribut privé de l'objet utilisateur le nombre de codes d'erreur reçus. Si ce nombre dépasse un seuil convenu, c'est que l'utilisateur connecté connaît mal le système. Une fonction inquisitrice est alors déclenchée demandant des renseignements supplémentaires à la personne connectée et interrompant la session en cas de réponses erronées.

### 5.5. l'inscription d'un utilisateur dans un groupe

Afin d'inscrire MICHEL sur la liste des membres de son groupe, l'objet "PAUL" envoie à l'objet MICHEL le message suivant :

| Mess ("PAUL","MICHEL",SIG_PAUL, inscription) |
|---|

L'objet MICHEL lui répond par un message d'acquittement contenant sa signature (normalement illisible) SIG_MICHEL, que l'objet PAUL va stocker dans sa liste de groupe.

| Mess("MICHEL","PAUL",SIG_MICHEL, ok) |
|---|

Remarques :

Si MICHEL est inscrit dans le groupe de PAUL, et que MICHEL fait rarement appel à l'objet "CIBLE" dont PAUL est le propriétaire, la perte de performance est tolérable. Par contre si MICHEL accède fréquemment à de nombreux objets de PAUL, la procédure de contrôle des accès devient lourde. MICHEL a la possibilité de persuader PAUL de lui donner des instances du type dont il a souvent besoin, ce qui le rendra propriétaire et annulera la procédure de contrôle. Toutefois PAUL peut refuser pour protéger la propriété intellectuelle sur ses objets.

## 6. LES CARACTERISTIQUES DE NOTRE APPROCHE

### 6.1. La représentation des utilisateurs dans le système
Chaque utilisateur fournit une représentation de lui même plus riche que dans la plupart des systèmes d'exploitation.

### 6.2. La performance
Le système de protection présenté est plus performant qu'un système hiérarchique classique car il est diffus, décentralisé, homogène, et systématique : la légalité de chaque message est contrôlé de manière transparente à l'utilisateur. En cas d'indiscrétion, (ce qui ne peut être empêché) l'intrus réussit à s'approprier les objets d'une personne, mais pas ceux des autres utilisateurs puisque leur protocole de reconnaissance est différent. L'encapsulation, les fonctions d'interfaces et les messages cristallisent la protection. Ce n'est qu'au moment de l'envoi d'un message à l'objet récepteur que le droit d'accès est évalué et communiqué à l'objet expéditeur et à lui seul. Les transactions demeurent secrètes grâce aux messages, dont la plupart des arguments sont implicites. Les messages représentent le seul moyen de connaître le contenu des objets, fournissant ainsi un moyen très sûr de protection. La lecture de la signature du propriétaire d'un objet n'apporte rien à un intrus car il ne peut produire de messages porteurs de cette signature. Personne ne peut connaître l'ensemble des liens de sécurité unissant les objets du système. Chaque utilisateur n'accède qu'à ses propres objets et à ceux pour lesquels un droit explicite et limité lui a été octroyé par son propriétaire.
D'un point de vue physique, les zones mémoires des objets sont allouées dynamiquement à des adresses variables, changeant à chaque connexion.

### 6.3. La souplesse
Nous proposons une organisation favorisant la création de petits groupes de travail autonomes qui convient bien à l'informatique scientifique. Tout se passe comme si chaque utilisateur était responsable de son système de protection.
Le protocole de reconnaissance est à la fois plus souple, plus mnémonique, simple ou complexe au gré de chacun. Le contrôle des messages est transparent pour l'utilisateur. Le système de protection est omniprésent, mais sa sévérité est modulée selon les situations : (vérification soigneuse des requêtes d'utilisateurs étrangers, déclenchement de fonctions inquisitrices, quand le taux d'erreurs trahit une méconnaissance du système). Une souplesse exceptionnelle est accordée au propriétaire qui peut : supprimer globalement les droits d'utilisation ou de lecture à un utilisateur en le retirant de son groupe, lui interdire quelques objets seulement, en modifiant leurs bits de protection, protéger la consultation de certains attributs jugés confidentiels, tout en laissant libre l'accès à d'autres.

### 6.4. L'aspect psychologique

Le système étant non-hiérarchisée, tous les utilisateurs ont les mêmes droits et les mêmes devoirs vis à vis des autres. Le dispositif de reconnaissance, étant librement choisi par chaque utilisateur, il est forcément bien accepté. Un chercheur ne peut s'épanouir que s'il est sûr de garder la maîtrise de sa production intellectuelle et que celle-ci ne sera pas pillée ou galvaudée. Il doit rester libre d'accorder le fruit de son travail aux seuls partenaires qu'il a choisis.

## 7. CONCLUSION

Les systèmes d'exploitation modernes devraient offrir un dispositif de protection suffisamment solide pour décourager la plupart des intrusions, tout en restant bien toléré par les utilisateurs. Les techniques d'effraction des systèmes informatiques restent nombreuses. Nous avons recensés les règles, qui gouvernent l'efficacité et l'acceptation d'un système de protection.

Nous proposons une approche orientée objet, souple, adaptable, respectant la volonté de chaque utilisateur de protéger ou de partager son travail avec autrui.

L'administrateur du système ne doit intervenir que pour introduire un nouvel utilisateur et dans les cas exceptionnels où des mesures de sauvegarde sont nécessaires.

Le dispositif de protection ne représente qu'un des nombreux problèmes à aborder dans le cadre de la réalisation d'un système d'exploitation. En ce moment même, le système UNIX est en train de bénéficier de l'apport des nouveaux langages orientés objet, notamment C++ [STRO 87][LEGG 88]. Nous sommes persuadés que les concepts des modèles objet et notamment l'encapsulation, les messages et les fonctions d'interfaces apporteront de nouvelles solutions en matière de protection.